\documentclass[12pt,preprint]{aastex}
\begin{document}
\title{BATC 13-band Photometry of Open Cluster NGC 7789}
\author{Zhen-Yu Wu, Xu Zhou, Jun Ma, Zhao-Ji Jiang, Jian-Sheng Chen, Jiang-Hua Wu}
\affil{National Astronomical Observatories, Chinese Academy of Sciences, 20A
Datun Road, Beijing 100012, China} \email{zywu@bac.pku.edu.cn}
\begin{abstract}
We present 13-band CCD intermediate-band spectrophotometry of a field centered
on the open cluster NGC 7789 from 400 to nearly 1000 nm, taken with
Beijing-Arizona-Taiwan-Connecticut (BATC) Multi-Color Survey photometric
system. By comparing observed spectral energy distributions of NGC 7789 stars
with theoretical ones, the fundamental parameters of this cluster are derived:
an age of $1.4\pm0.1$ Gyr, a distance modulus $(m-M)_{0}=11.27\pm0.04$, a
reddening $E(B-V)=0.28\pm0.02$, and a metallicity with the solar composition
$Z=0.019$. When the surface density profile for member stars with limiting
magnitudes of 19.0 in the BATC $e$ band ($\lambda_{\textrm{eff}}=4925$ \AA) is
fitted by King model, the core radius $R_{c}=7.52\arcmin$ and the tidal radius
$R_{t}=28.84\arcmin$ are derived for NGC 7789. The observed mass function (MF)
for main sequence stars of NGC 7789 with masses from 0.95 to 1.85 $M_{\odot}$
is fitted with a power-law function $\phi(m)\propto m^{\alpha}$ and the slope
$\alpha=-0.96$ is derived. Strong mass segregation in NGC 7789 is reflected in
the significant variation of the concentration parameters
$C_{0}=\log\,(R_{t}/R_{c})$ for member stars of NGC 7789 within different mass
ranges: $C_{0}=1.02$ for most of massive stars; $C_{0}=0.37$ for the
lowest-mass MS stars. Strong mass segregation in NGC 7789 is also indicated in
the significant variation of the slopes $\alpha$ in different spatial regions
of the cluster: the MF for stars within the core region has $\alpha=-0.71$,
much flatter than that for stars in external regions of the cluster
($\alpha=-1.20$).
\end{abstract}
\section{INTRODUCTION}
Open clusters (OCs) have long been recognized as important tools in the study
of stellar formation and evolution. It is becoming increasingly clear that most
stars do not form in isolation but are instead born in relatively rich
clusters. These clusters range from small associations of some tens of stars to
OCs of several hundreds of stars and very rich dense OCs containing several
thousands or more stars \citep[and references therein]{bd98}.

Cluster evolution is an intricate mix of dynamics, stellar evolution and
external tidal influences \citep{pz01,be01}. Generally speaking, mass-loss from
stellar evolution is of greatest importance during the first few tens of
millions of years of cluster evolution, and may well result in the disruption
of the entire cluster. If the cluster survives this early phase, stellar
evolutionary time-scales soon become longer than the time-scales for dynamical
evolution, and two-body relaxation and tidal effects become dominant.
Ultimately, these effects cause the cluster to dissociate and the stars to
become part of the general field population of the Galactic disc \citep{pz01}.

Energy equipartition in the cluster is the main dynamical result of tow body
relaxation and leads to the lower mass stars move outwards and the higher mass
stars inward, i.e., mass segregation. Numerical simulations have shown that
dynamical mass segregation occurs on approximately the cluster relaxation time.
In a relaxation time the cluster develops a well-defined dual regions: the core
and the halo \citep{de97,bd98}. Mass segregation combined with the stripping of
stars by the tidal field of the Galaxy, alters the mass function (MF) of the
cluster over time to give a deficiency of low-mass stars and flattens the MF
slope of the cluster \citep{de97,kr01,hu05}. As the result of dynamical
evolution of the cluster over time, mass segregation has been observed in many
intermediate-age and old open clusters \citep{bb05}.

The Beijing-Arizona-Taiwan-Connecticut (BATC) Multi-Color Survey photometric
system consists of 15 filters of band-widths 150 --- 350 \AA, which cover the
wavelength range 3300 --- 10000 \AA. This photometric system is designed to
avoid strong and variable sky emission lines \citep{fan}, and uses a CCD with a
field of view $58\arcmin\times58\arcmin$. The observed field size and
photometric depth of BATC photometric system make it suitable for studying OCs
in the Galaxy. This photometric system has been used to derive fundamental
parameters and membership of the open cluster M48 \citep{wu05,wu06} and to
study mass segregation and MF in the old open cluster M67 \citep{fan}. In this
paper we will derive the fundamental parameters of the intermediate-age open
cluster NGC 7789 and study the dynamical states and the MF of this cluster
based on the observed data taken with the BATC photometric system.

The open cluster NGC 7789 (C 2354+564)
($\alpha_{2000}=23^{\textrm{h}}57^{\textrm{m}}24^{\textrm{s}}$,
$\delta_{2000}=+56\arcdeg 42\arcmin 30\arcsec$; $l=115\fdg532$, $b=-5\fdg385$)
has been studied with numerous photometric and spectral observations because of
its very rich stellar population (e.g., Gim et al. 1998b;
Barta\v{s}i\={u}t\.{e} \& Tautvai\v{s}ien\.{e} 2004). The first extensive $UBV$
photoelectrical and photographic photometry of this cluster was obtained by
\citet{bu58}. Their color-magnitude diagram (CMD) shows a well-defined and
extended red giant branch (RGB), a prominent ``clump'' of core He-burning
stars, numerous blue stragglers, and a main sequence (MS) whose upper end bends
significantly to the red. After the study of NGC 7789 by \citet{bu58},
individual giant stars and blue stragglers in this cluster have been used to
derive the basic parameters such as reddening, distance modulus, and
metallicity through various photometric ($UBV$, $uvby$, DDO, and Washington)
and spectroscopic observations. The extensive CCD photometric observations have
been obtained in recent years in $BV$ \citep{ma94}, $VI$ \citep{gim98b}, and
$JK$ \citep{va00} filters, and the age of this cluster has been derived from
those photometric data. In Table \ref{told}, we summarize the available
fundamental parameters of NGC 7789 from literature. Table \ref{told} shows that
the derived fundamental parameters of NGC 7789 lie in a quite wide range: the
reddening $E(B-V)$ from 0.22 to 0.35 with a mean of $0.28 \pm 0.01$; the
distance modulus $(m-M)_{0}$ from 11.0 to 11.7 with a mean of $11.38 \pm 0.05$;
the metallicity [Fe/H] from $-0.35$ to the value of solar with a mean of
$-0.14\pm0.03$; and the age from 1.1 Gyr to 1.7 Gyr with a mean of $1.37 \pm
0.1$ Gyr. \citet{mc81} have determined relative proper motions of 1387 stars in
the field of NGC 7789, 679 of which are probable members brighter than
$B\approx15.5$. Radial velocity observations have been done by \citet[and
references therein]{fr02} and \citet{gim98a}.

In this paper, the fundamental parameters and membership of open cluster NGC
7789 are derived with the new method based on comparison of the spectral energy
distributions (SEDs) of stars with the theoretical ones \citep{wu05,wu06}.
Using the candidate member stars determined by our photometric method, the mass
segregation and MF of NGC 7789 are discussed in detail.

In the following, we describe our new observational data and photometric
reduction of NGC 7789 in Section 2. In Section 3, we derive the fundamental
parameters of NGC 7789. The observed MF of stars in NGC 7789 and the phenomenon
of mass segregation in this cluster are discussed in Section 4. And a summary
are presented in Section 5.
\section{OBSERVATIONS AND  DATA REDUCTION}
\subsection{Observations}
\begin{figure}
%\epsscale{1.0}
\plotone{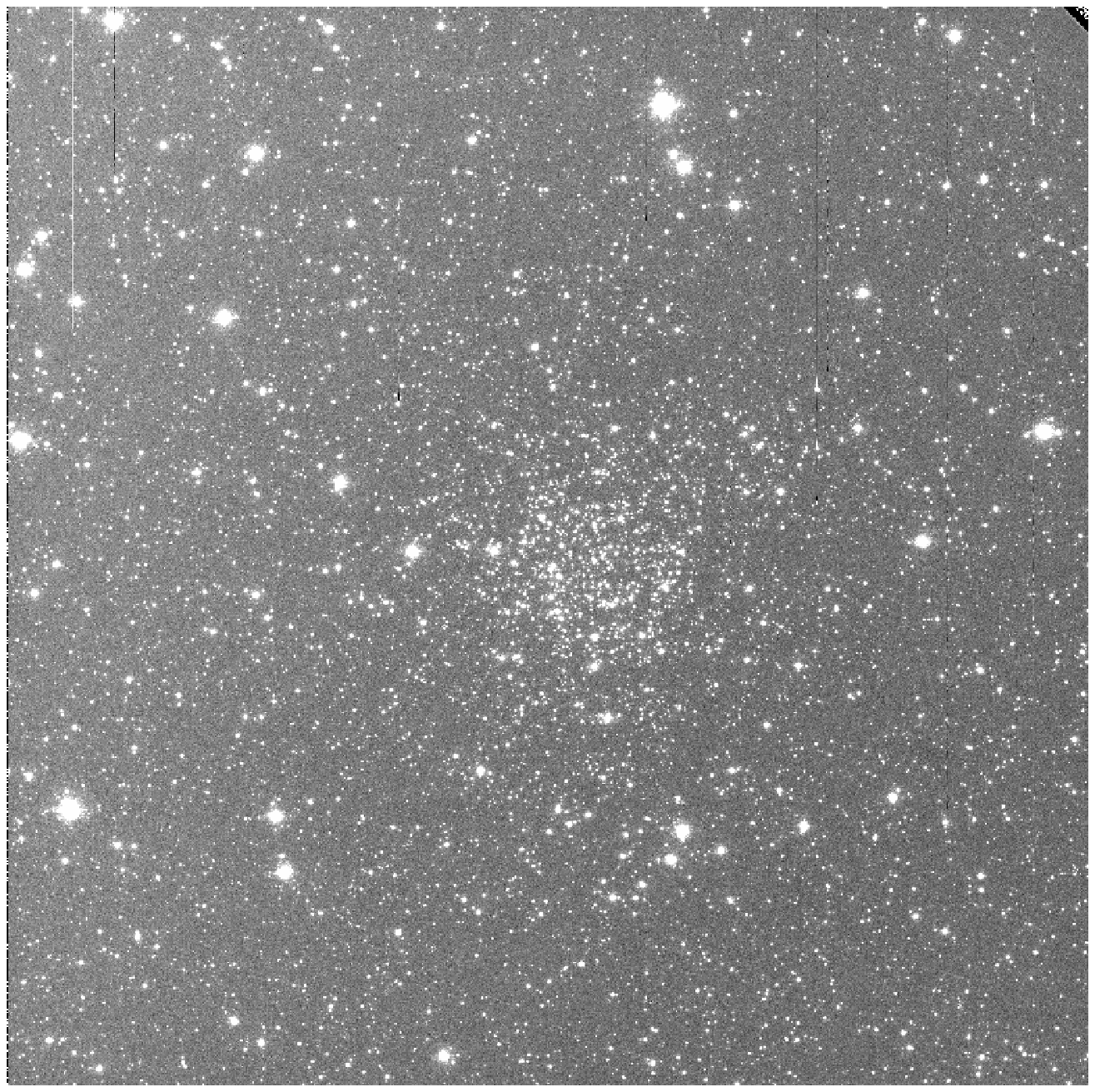} \caption{The exposed 600s image in BATC $e$ band centered on
NGC 7789 with a view of field of $58\arcmin\times58\arcmin$.\label{fxy}}
\end{figure}

Our observations were conducted with the BATC photometric system at the
Xinglong Station of the National Astronomical Observatories, Chinese Academy of
Sciences (NAOC). The 60/90 cm f/3 Schmidt telescope was used, with a Ford
Aerospace $2048\times2048$ CCD camera at its main focus. The field of view of
the CCD is $58\arcmin\times58\arcmin$, with a scale of $1\farcs7$ pixel$^{-1}$.
A image in the BATC $e$ band, which was exposed 600 seconds, is presented in
Figure \ref{fxy}.

The filter system of the BATC project is defined by 15 intermediate-band
filters that are specifically designed to avoid most of the known bright and
variable night-sky emission lines. The definition of magnitude for the BATC
survey and the observing procedure of the survey program field and photometry
are described in detail in \citet{yan} and \citet{zhou01}.

Because of the very low quantum efficiency of the thick CCD used in the bluest
filters, two BATC filters ($a$ and $b$) are not used in the observation of the
NGC 7789 field. In Table \ref{tobs}, for each BATC filter, we list the
corresponding effective wavelength, FWHM, exposure time, and the number of
frames observed.
\subsection{Data Reduction and Calibration}
Preliminary reductions of  the CCD frames, including bias subtraction and field
flattening, were carried out with an automatic data reduction procedure called
PIPELINE I, which has been developed as a standard for the BATC survey in NAOC
\citep{fan}. A PIPELINE II program based on the DAOPHOT II stellar photometric
reduction package of \citet{stet} was used to measure the instrumental
magnitudes of point sources in BATC CCD frames. The PIPELINE II reduction
procedure was performed on each single CCD frame to get the point-spread
function (PSF) magnitude of each point source and then the aperture correction
was performed to get the total instrumental magnitude for each star. The
instrumental magnitudes were then calibrated to the BATC standard system
\citep{zhou03}. The average calibration error of each filter is less than 0.02
mag. The mean of FWHM of the PSF in each observed image is 4.39$\arcsec$ with
1$\sigma$ error of 0.79$\arcsec$, which critically samples ($\sim$ 2.5 pixels)
the PSF. Therefore, the effect of undersampling on our photometric accuracy is
not critical \citep{fan}. For each star observed more than once in a BATC band,
the final photometric result in that band is the weighted mean. Stars that are
detected in at least six bands are included in the final star catalogue.
\subsection{Artificial Star Test}
In order to obtain the MF of NGC 7789, we estimated the completeness of our
sample. Completeness corrections have been determined by artificial stars test
on our data. We created 5 artificial images for each observed image by adding
to the original images artificial stars. In order to avoid the creation of
overcrowding, in each test we added at random positions only 15\% of the
original number of stars. The artificial stars had the same PSF and luminosity
distribution of the original sample. The test images were reduced following the
same steps and using the same parameters as on the original images. The
completeness corrections for the whole image and for two different regions of
the cluster, defined as the ratio of the found stars over the added artificial
stars are listed in Table \ref{tadds} for the BATC $e$ magnitude. In the MFs we
included only the values for which the completeness corrections were 0.5 or
higher for all regions in this cluster. Therefore we set the limiting magnitude
to $e=19.0$.

\section{FUNDAMENTAL PARAMETERS DETERMINATION}
\subsection{Theoretical Model and Stars Used}
\begin{figure}
 \plotone{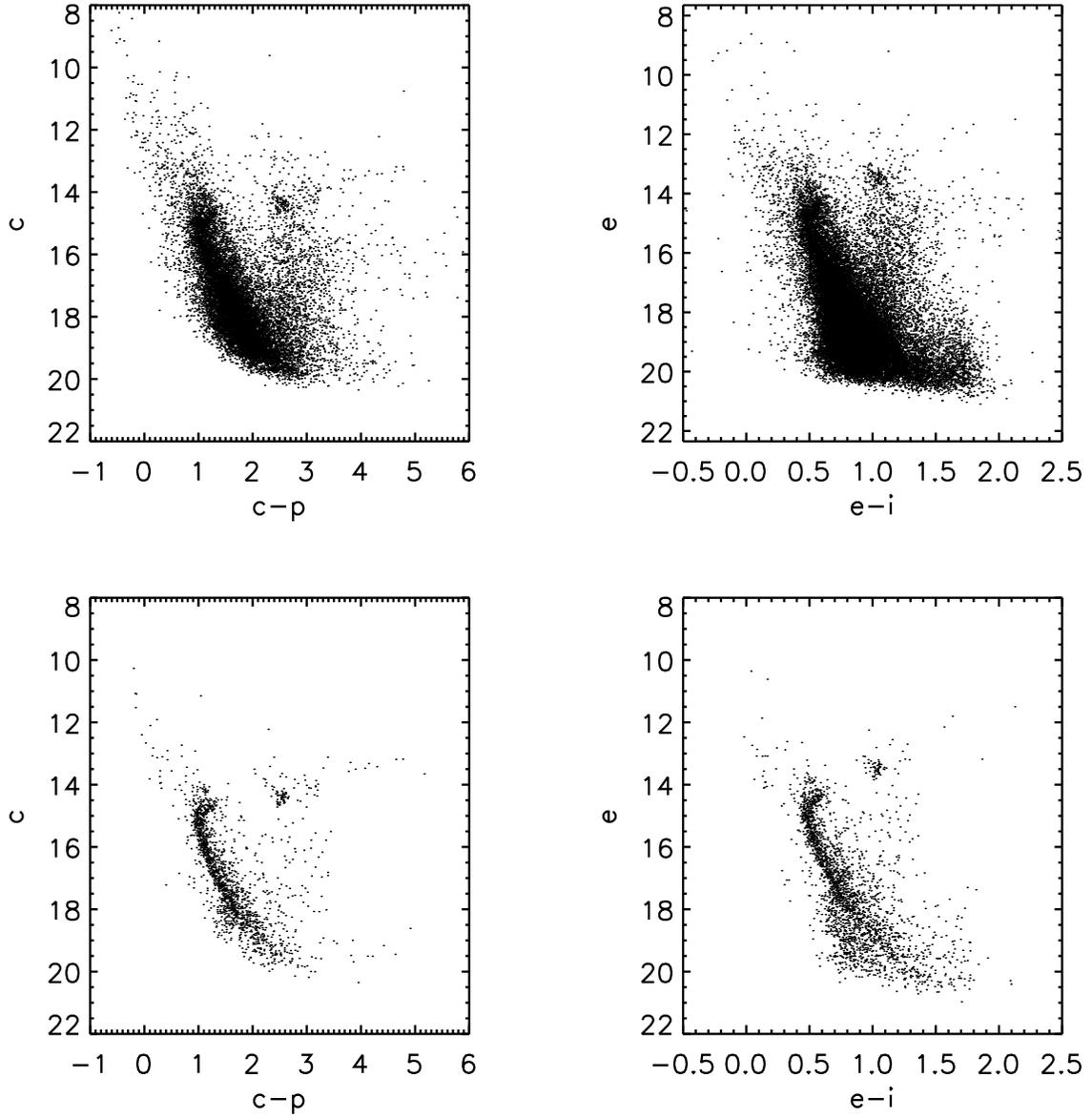}
\caption{The BATC ($c-p$) vs. $c$ and ($e-i$) vs. $e$ CMDs for NGC 7789. The
top panels are for all stars observed around NGC 7789; the bottom panels are
for stars within $8\arcmin$ (the core radius) of the cluster
center.\label{fcmd}}
\end{figure}

The top two panels of Figure \ref{fcmd} show the CMDs of NGC 7789 for all
observed stars around the cluster in BATC ($c-p$) vs. $c$ and ($e-i$) vs. $e$
bands. The bottom two panels of Figure \ref{fcmd} only show CMDs for stars
within $8\arcmin$ (the core radius, see next section) of the cluster center. It
can be seen that our photometry extends from near the RGB tip to as faint as
$\sim$ 5 mag below the MS turnoff. The CMDs also show a prominent ``clump", and
numerous blue straggler candidates. The bottom panels of Figure \ref{fcmd} much
clearly show the MS of this cluster.

To apply our method to derive the fundamental parameters for NGC 7789, stars
with BATC $e$ magnitudes less than 18.0  and within $5\arcmin$ of the cluster
center are chosen as our input sample.  The radius of $5\arcmin$ is set to
discard most of contamination by field stars. The magnitude is set to eliminate
the effect of large photometric uncertainties at faint magnitudes in our data.
1142 stars are included in our last sample.

Padova stellar evolutionary models \citep[and references therein ; hereafter
``Padova 2000"]{gir00,gir02} are used in our present method. Padova 2000 models
present a large grid of stellar evolutionary tracks and isochrones that are
suitable for modeling star clusters.  These models are computed with updated
opacities and equation of state, and also a moderate amount of convective
overshoot. They also present an additional set of models with solar
composition, computed using the classical Schwarzschild criterion for
convective boundaries(i.e., without overshoot)\footnote{The Padova stellar
evolutionary models in BATC photometric system can be downloaded at
``http://pleiadi.pd.astro.it/isoc\_photsys.02/isoc\_batc/index.html''.}.

\subsection{Method}
In the present work, 13 bands data are obtained for stars in NGC 7789, which
provide a sort of \textsl{low-resolution spectroscopy} that defines the SED for
each star. Following the method presented in our previous papers
\citep{wu05,wu06}, the fundamental parameters: metallicity, age, reddening and
distance modulus are derived by fitting the observed SEDs of member stars in
NGC 7789 with the theoretical ones of Padova 2000 models. For the $j$th star, a
parameter $S$ can be defined:
\begin{equation}
S_{j}[t,Z,(m-M)_{0},E(B-V)]=\sum_{i=1}^n\frac{\{m_{ij}-M_{i}[t,Z,(m-M)_{0},E(B-V)
]\}^{2 } } { \sigma_ { ij}^{2}}
\end{equation}
where $M_{i}[t,Z,(m-M)_{0},E(B-V)]$ is the theoretical magnitude in the $i$th
BATC band, corrected by distance modulus $(m-M)_{0}$ and reddening $E(B-V)$ and
computed from the chosen theoretical isochrone model with age $t$, metallicity
$Z$. The reddening $E(B-V)$ is transformed to each BATC band using the
extinction coefficient derived by \citet{ch00}, based on the procedure given in
Appendix B of \citet{sch}. Here $m_{ij}$ and $\sigma_{ij}$ are the observed
magnitude and its error, respectively, of the $j$th star in the $i$th band, and
$n$ is the total number of observed bands for the $j$th star. For $M_{i}$ with
different stellar masses, the minimum of $S_{j}$, $S_{j,\textrm{min}}$ can be
obtained for the $j$th star with the chosen theoretical models. If the observed
SEDs can match the theoretical SEDs, the parameter $S_{\textrm{min}}$ should be
the $\chi^{2}$ distribution with $n-P$ degrees of freedom, where $P$ is the
number of free parameters to be solved. The integral probability at least as
large as $S_{j,\textrm{min}}$ in the $\chi^{2}$ distribution with $n-P$ degrees
of freedom is taken as the ``photometric'' membership probability of the $j$th
star \citep{wu06}. For the given stellar sample, the parameter
$S_{\textrm{min}}$ is calculated for each star. For stars with photometric
membership probabilities $P_{\textrm{\small phot}}$ greater than 10\%, another
parameter can be defined:
\begin{equation}
S_{c}[t,Z,(m-M)_{0},E(B-V)]=\frac{\sum_{j=1}^{N_{\textrm{\tiny{mem}}}}S_{j,\textrm{min}}
}{N_{\textrm{mem}}}
\end{equation}
where $N_{\textrm{mem}}$ is the number of ``photometric'' member stars with
$P_{\textrm{\small phot}}$ greater than 10\%. We calculate the parameter
$S_{c}$ for various parametric sets with different distance modulus, reddening,
age and metallicity. The parametric set with the maximum $N_{\textrm{mem}}$ and
minimum $S_{c}$ is considered as the best-fitting one for this cluster
\citep{wu06}. In order to apply the above method to NGC 7789, the parameters
are chosen as follows: metallicity are taken to be $Z =
0.004,0.008,0.019,0.03$; ages $\log\,t$ from 9.0 to 9.4, in steps of 0.05;
distance modulus $(m-M)_{0}$ from 11.1 to 11.5, in steps of 0.01; and reddening
$E(B-V)$ from 0.2 to 0.4, in steps of 0.01.

\subsection{RESULTS AND DISCUSSION}
\subsubsection{Derived Fundamental Parameters}
\begin{figure}
 \plotone{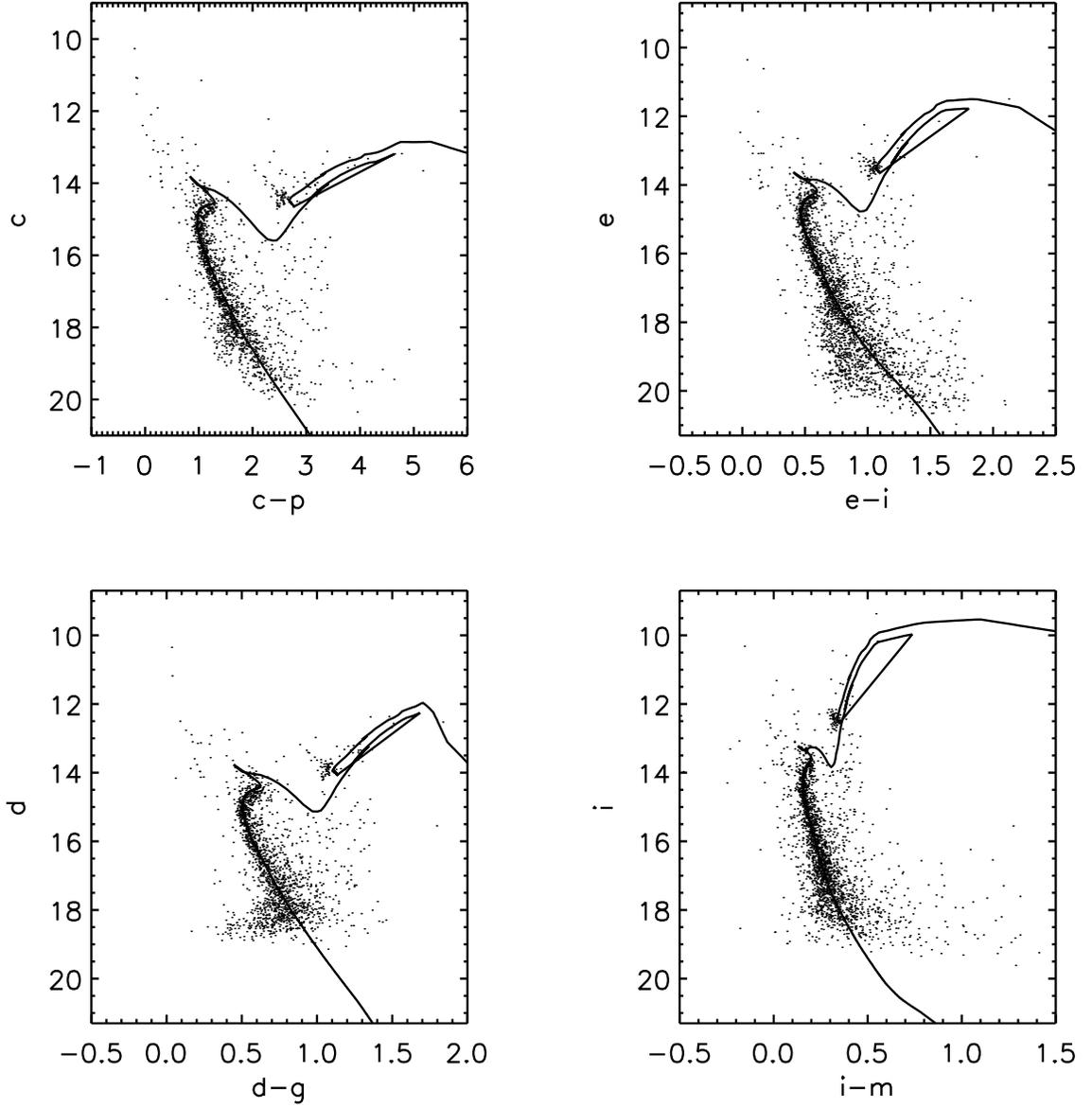}
\caption{The BATC ($c-p$) vs. $c$, ($e-i$) vs. $e$, ($d-g$) vs. $d$ and ($i-m$)
vs. $i$ CMDs of NGC 7789 for stars within $8\arcmin$ of the cluster center,
along with the best-fitting Padova 2000 theoretical isochrones computed with a
moderate amount of convective overshoot. The derived best-fitting parameters
for NGC 7789 are adopted as follows: metallicity $Z=0.019$, age $\log\,t=9.15$,
distance modulus $(m-M)_{0}=11.27$, and reddening $E(B-V)=0.28$. \label{fiso}}
\end{figure}

Applying the above method, we find that with a distance modulus
$(m-M)_{0}=11.27$ and a reddening $E(B-V)=0.28$, the theoretical model with an
age of $\log\,t=9.15$ (1.4 Gyr) and metallicity $Z=0.019$ can best fit the
observed SEDs of member stars in NGC 7789. 822 stars are determined as
photometric members with $P_{\textrm{\small phot}}$ greater than 10\%. In
Figure \ref{fiso}, using the derived best-fitting parameters for NGC 7789, the
theoretical isochrones are plotted in the BATC ($c-p$) vs. $c$, ($e-i$) vs.
$e$, ($d-g$) vs. $d$ and ($i-m$) vs. $i$ CMDs of NGC 7789 for stars within
$8\arcmin$ of the cluster center. Figure \ref{fiso} indicates that adopting our
derived best-fitting parameters for NGC 7789, the theoretical isochrones
reproduce the observed cluster's MS, RGB fiducial and red clump stars.

\begin{figure}
 \plotone{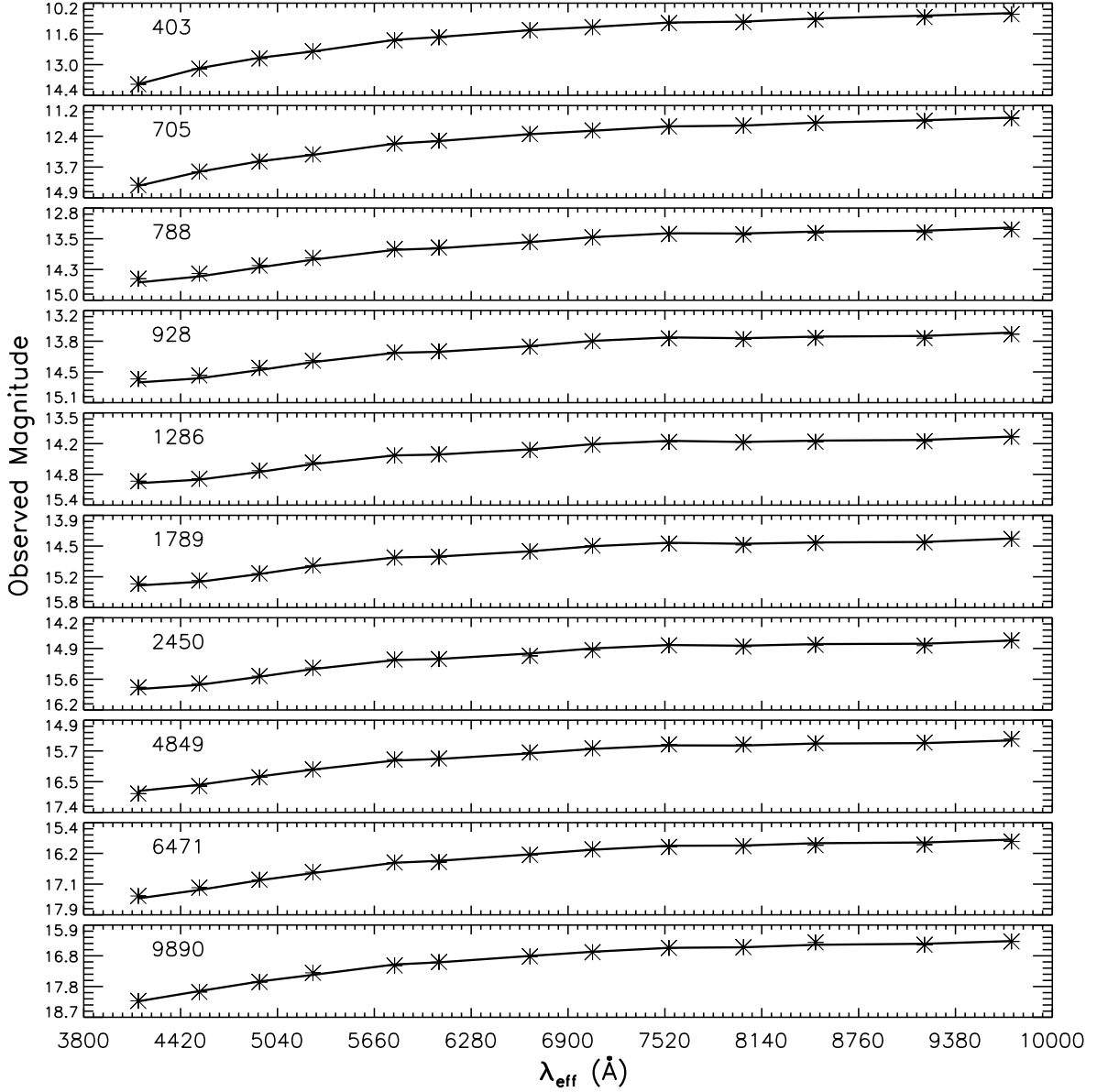}
\caption{SEDs of member stars in NGC 7789. The asterisks are for observed SEDs,
the solid lines are for theoretical SEDs with our derived best-fitting
parameters for this cluster:  Metallicity $Z=0.019$, age $\log\,t=9.15$,
distance modulus $(m-M)_{0}=11.27$, and reddening $E(B-V)=0.28$. The
identification number (ID) for each star in our star catalog is labeled in each
panel. \label{fsed}}
\end{figure}

Figure \ref{fsed} shows the SEDs for a sample of 10 RGB and MS stars in NGC
7789 with $P_{\textrm{\small phot}}$ greater than 90\%. The asterisks are for
observed SEDs and the solid lines are for theoretical SEDs. Figure \ref{fsed}
indicates that adopting our derived best-fitting parameters for NGC 7789, the
theoretical SEDs can fit the observed SEDs very well.

11 member blue stragglers of NGC 7789 that identified by \cite{gim98b} were
included in our sample. Using our derived best-fitting parameters for NGC 7789,
there are no theoretical SEDs can fit the observed SEDs of these blue
stragglers. In order to find the best-fitting theoretical SEDs for these blue
stragglers, we use the derived metallicity, distance modulus and reddening for
this cluster and change the age of theoretical isochrones from $\log\,t=8.0$ to
$\log\,t=9.4$. We find that the theoretical SEDs with an age of $\log\,t=8.75$
($t=0.56$ Gyr) can fit the observed ones of these blue stragglers. Thus, the
observed SEDs of blue stragglers can be fitted by more younger stellar models
than those that matched the other stars in the same cluster. The mean mass
derived for these blue stragglers is $2.2\,M_{\odot}$, much bigger than that
for the turnoff stars in the cluster($1.8 \,M_{\odot}$).

Using our derived best-fitting parameters for NGC 7789, the membership
probabilities of all observed stars are determined by comparing observed SEDs
with theoretical ones and a 10\% limiting probability is set to distinguish
cluster and field stars.

\subsubsection{The Effects of Different Limiting Membership Probabilities and Theoretical Models}
To check the effect of the chosen limiting probability for our derived
parameters, we repeated the above fitting procedure with other two limiting
probabilities: 50\% and 1\%. For the 50\% limiting probability, the derived
best-fitting parameters for NGC 7789 are as follows: metallicity $Z=0.019$, age
$\log\,t=9.15$, reddening $E(B-V)=0.28$, and distance modulus
$(m-M)_{0}=11.30$. 728 stars are considered as member stars with the 50\%
limiting probability. For the 1\% limiting probability, the derived
best-fitting parameters are as follows: metallicity $Z=0.019$, age
$\log\,t=9.10$, reddening $E(B-V)=0.30$, and distance modulus
$(m-M)_{0}=11.32$. 881 stars are considered as member stars with the 1\%
limiting probability. It is clear that the derived metallicity, age, reddening
and distance modulus are consistent for all the adopted different limiting
probabilities.

We also chose 348 stars with proper-motion-based probabilities greater than
80\% and with $P_{\textrm{\small phot}}$ greater than 90\% as member stars, the
derived best-fitting parameters for NGC 7789 are as follows: an age
$\log\,t=9.15$, reddening $E(B-V)=0.28$, $(m-M)_0=11.39$, and $Z=0.019$. This
result is consistent with that derived for member stars only with
$P_{\textrm{\small phot}}$ greater than 10\%. The internal uncertainties for
our derived best-fitting parameters are as follows: $\log\,t=\pm0.05$
($t=\pm0.1$ Gyr), $(m-M)_{0}=\pm0.04$, and reddening $E(B-V)=\pm0.02$, which
include the influence of observational errors and the estimated uncertainty
inherent in our fitting method. The derived metallicity is not affected by
above uncertainties, which indicates a very limited sensibility of metallicity
to the variations in the limiting probabilities.

\begin{figure}
 \plotone{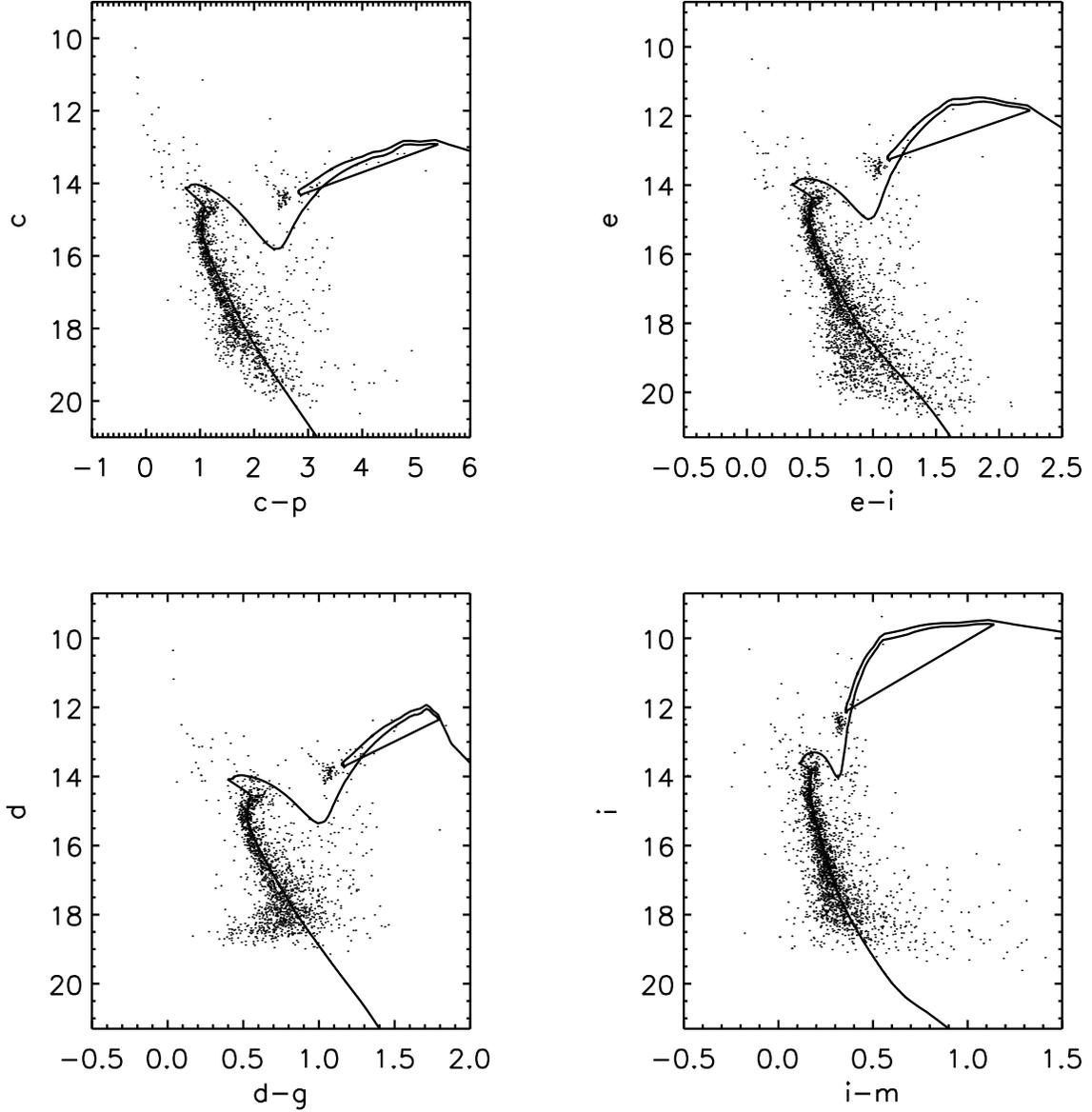}
\caption{Same as Fig \ref{fiso}, but the best-fitting Padova 2000 theoretical
isochrones computed with nonovershoot and the fundamental parameters for NGC
7789 are adopted as follows: Metallicity $Z=0.019$, age $\log\,t=9.05$,
distance modulus $(m-M)_{0}=11.10$, and reddening $E(B-V)=0.29$. Comparing with
the result presented in Fig \ref{fiso} derived with overshoot theoretical
model, the best-fitting theoretical isochrone with nonovershoot can not
reproduce the observed cluster's MS turnoff, RGB fiducial and red clump
stars.\label{fiso2}}
\end{figure}

As an intermediate-age open cluster, NGC 7789 is a good object to test the
stellar evolutionary models computed with or without convective overshoot.
Using the Padova 2000 theoretical model computed with nonovershoot and solar
composition, we derive the best-fitting parameters for NGC 7789 as follows: age
$\log\,t=9.05$ (1.1 Gyr), distance modulus $(m-M)_{0}=11.10$, and reddening
$E(B-V)=0.29$. In Figure \ref{fiso2}, using the above derived fundamental
parameters, we overimposed the theoretical isochrones computed with
nonovershoot to the CMDs of NGC 7789 in BATC ($c-p$) vs. $c$, ($e-i$) vs. $e$,
($d-g$) vs. $d$ and ($i-m$) vs. $i$ bands. Figure \ref{fiso2} indicates that
theoretical models computed with nonovershoot can not reproduce the observed
cluster's  MS turnoff, RGB fiducial and red clump stars in CMDs of NGC 7789.

\subsubsection{Comparison with Previous Studies}
In this section, we compare our derived best-fitting parameters for NGC 7789
with those listed in Table \ref{told}. For metallicity, the theoretical model
computed with solar composition $Z=0.019$ match our observed data very well.
This metallicity is very close to the recently derived results
$\textrm{[Fe/H]}=-0.04$ given by \citet{ta05} who obtained high-resolution CCD
spectroscopy of 9 RGB stars in NGC 7789, and also close to the photometric
result of \citet{tw97}. Our result is also consistent, within the errors, with
that derived by \citet{pi85} based on high-resolution CCD spectroscopy of 6 RGB
stars in the range of quoted errors.  For reddening $E(B-V)$, our derived
best-fitting result of 0.28 is same as the mean listed in Table \ref{told}. For
distance modulus $(m-M)_{0}$, our derived value of $11.27$ is less than the
mean of 11.38 listed in Table \ref{told}. Our derived age of 1.4 Gyr for NGC
7789 is consistent with recent results of \citet{gim98b} and \citet{va00},
which are both based on fitting isochrones models computed with overshoot.

590 stars in our sample have proper-motion-based membership probabilities
determined by \cite{mc81}. Adopting the same limiting probability 10\% for
cluster-field segregation to proper-motion-based membership probabilities, we
can get a 73\% agreement between the present method and proper-motion-based
method, which is close to the value 80\% derived for open cluster M48 in our
previous paper \citep{wu06}.

\section{MASS SEGREGATION AND MASS FUNCTION}
\subsection{Mass Segregation}
In order to study mass segregation in NGC 7789, we examine the radial surface
density profile of candidate cluster members of NGC 7789. Stars within
$25\arcmin$ from the cluster center with limiting magnitudes of 19.0 in BATC
$e$ band and $P_{\textrm{\small phot}}$ greater than 10\% are chosen as
preliminary members. 11 member blue stragglers identified by \citet{gim98b} are
also included in our sample. Our candidate member stars in NGC 7789 include
blue stragglers, RGB stars, subgiant branch stars, stars near the MS turnoff
and MS stars extending to $\sim19.0$ in the BATC $e$ band. The masses of stars
extend from $\sim 0.8 $ to $\sim2.0\,M_{\odot}$.

\begin{figure}
 \plotone{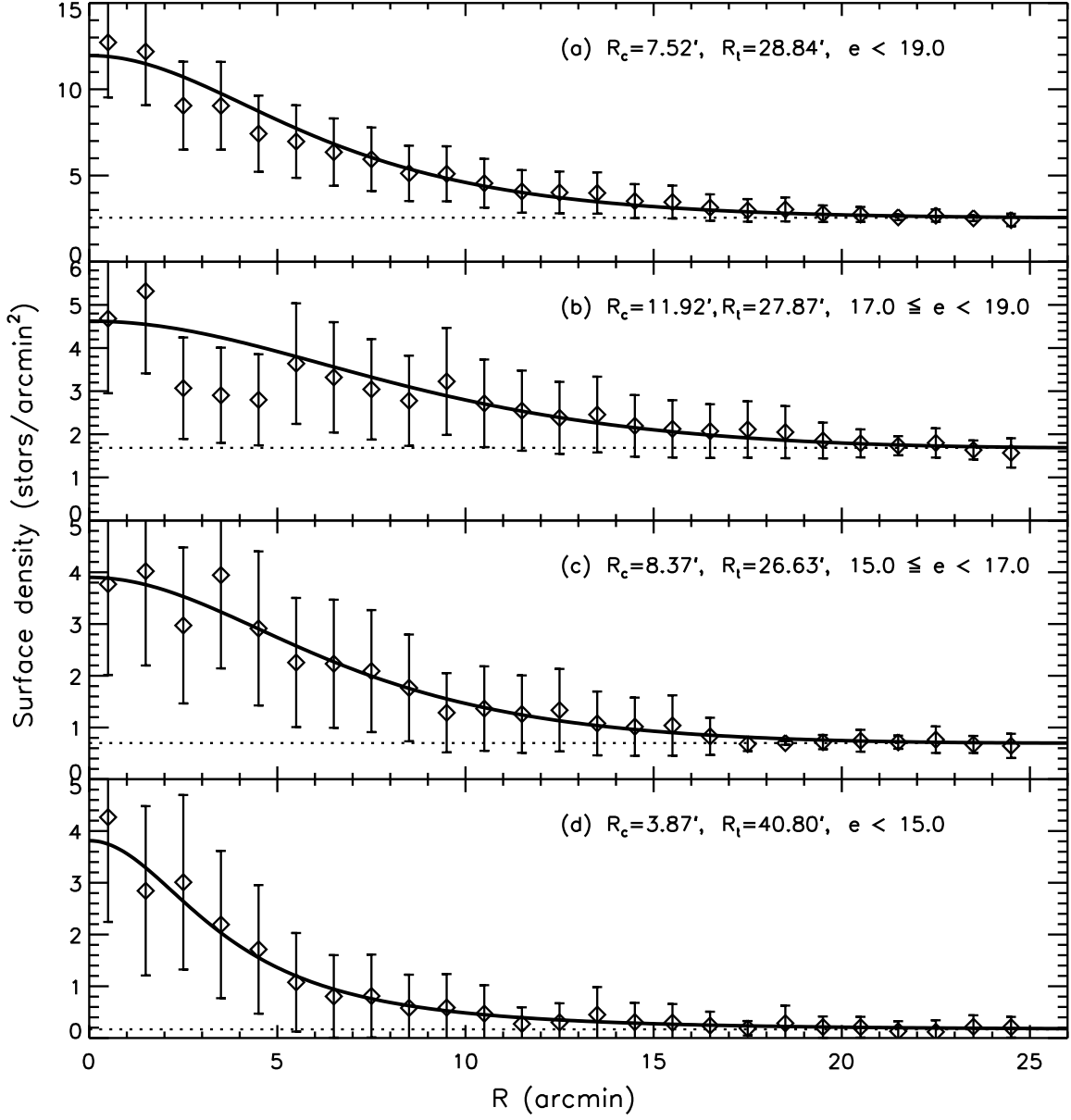}
\caption{Radial surface density profile of member stars in NGC 7789 with
photometric membership probabilities greater than 10\%. The average background
levels are shown as dotted lines and error bars present $1\,\sigma$ Poisson
errors. The solid lines show the best-fitting King models to the radial
profiles. The fitted King model parameters: core radii $R_{c}$ and tidal radii
$R_{t}$ are labeled in each panel for the labeled different magnitude
groups.\label{fking}}
\end{figure}

To determine the radial surface density, the cluster was divided into a number
of concentric rings with a step of $1.0\arcmin$. The radial stellar density in
each concentric circle was obtained by dividing the number of stars in each
annulus by its area. To check mass segregation in NGC 7789, candidate member
stars were placed in three magnitude (mass) groups based on their BATC $e$
magnitude: stars within $17.0\leq e<19.0$, within $15.0\leq e<17.0$ and within
$e<15.0$, respectively. Figure \ref{fking} shows the radial surface density
profiles for different magnitude groups. Panel a of Figure \ref{fking} shows
radial surface density profile for all candidate member stars within $e<19.0$,
in the region of $21 \arcmin \leq R \leq 25\arcmin$, the stellar density
distribution is flat and nonzero, which indicates the contamination of field
stars in our preliminary candidate member stars. For different magnitude
groups, the field-star contamination corresponding to the average number of
stars included in the rings located at $21 \arcmin \leq R \leq 25\arcmin$ is
represented in Figure \ref{fking} and indicated by dotted lines.

We fitted the empirical density law of \citet{king62} to our field-subtracted
radial density profiles for different magnitude groups of candidate member
stars in NGC 7789. The fit was performed using a nonlinear least squares fit
routine which used the $1\,\sigma$ Poisson errors as weights.  The fitted King
model parameters core radius $R_{c}$ and tidal radius $R_{t}$ are labeled in
each panel of Figure \ref{fking} for different magnitude groups. The fitting
results are represented with solid lines. Figure \ref{fking} shows that within
uncertainties the King model reproduces well the radial surface density
profiles for all magnitude groups. Figure \ref{fking} indicates that evidence
for mass segregation is strong when comparing King model fitting results for
different magnitude groups: the group with BATC $e<15.0$ including most massive
stars such as blue straggler, RGB stars, subgiant branch stars and MS turnoff
stars, has the smallest core radius and the largest tidal radius; the group
with BATC $17.0 \leq e<19.0$ including most low-mass MS stars, has the largest
core radius about three times than that for high-mass stars. To represent the
mass segregation more clearly, the concentration parameter
$C_{0}=\log\,(R_{t}/R_{c})$ \citep{king62} is derived for stars within
different magnitude groups and listed in Table \ref{tking}. The derived
concentration parameters decrease from the high-mass group to the low-mass one
and clearly indicate that high-mass stars are more centrally concentrated than
low-mass stars.

In Table \ref{tking}, we also present the radial surface density profiles for
stars with $P_{\textrm{\small phot}}$ greater than 50\% and 1\% respectively.
For different magnitude groups, the derived King model parameters are
consistent within uncertainties with that derived with limiting probability of
10\%. The concentration parameters are also derived and listed in Table
\ref{tking}. Table \ref{tking} shows that there is strong evidence for mass
segregation in NGC 7789 based on surface density profiles for different
magnitude (mass) groups. Adopting different limiting probabilities for
photometric member stars doesn't affect the mass segregation represented in the
present data.

\subsection{Mass Functions}
The MF of stars can be fitted with a power-law function $\phi(m) \propto
m^{\alpha}$ \citep{kr01}. The present observation, which covers a large field
of view of NGC 7789, provides an opportunity to study the spatial distribution
of MF in this cluster. The procedure to choose preliminary candidate member
stars is the same as in the previous section, but we take the MS turnoff as the
bright limit ($e \sim 13.5$). The mass of each candidate member star is
determined in the procedure of fitting observed SED of member star with the
theoretical model, in which our derived best-fitting parameters for NGC 7789
are adopted. The final sample includes member stars with masses from $\sim 0.8$
to $\sim 1.85 \,M_{\odot}$.

\begin{figure}
 \plotone{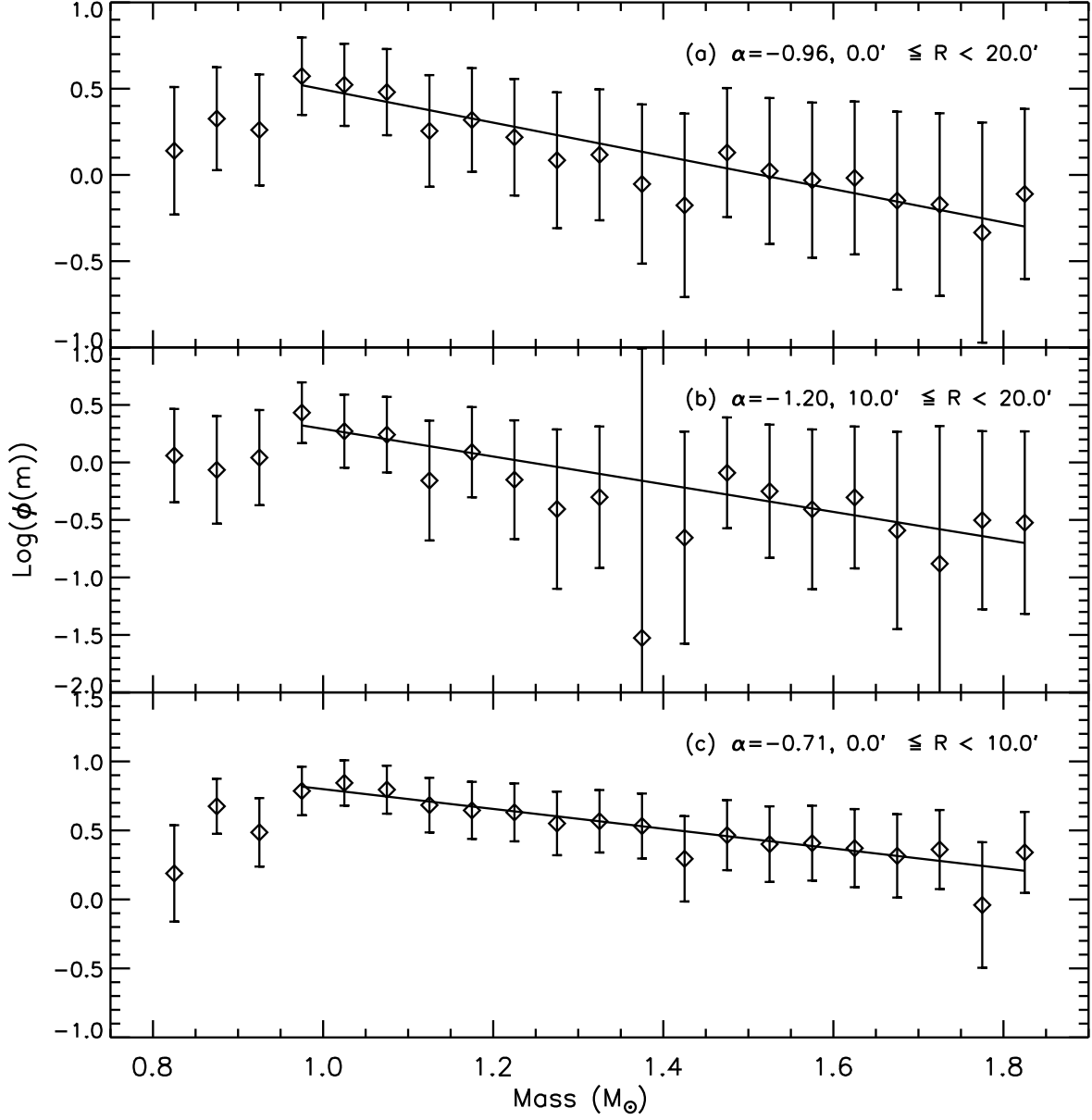}
\caption{Field-corrected mass functions in different spatial regions of NGC
7789. Error bars represent the $1\,\sigma$ Poisson errors. Solid lines indicate
the best-fitting power-law function $\phi(m)\propto m^{\alpha}$ for stars with
mass between 0.95 and 1.85 $M_{\odot}$.\label{fmf}}
\end{figure}

After having corrected for completeness in different regions of NGC 7789, the
MFs for different regions of this cluster are presented in Figure \ref{fmf}.
The field-star contamination is estimated in the region at $21\arcmin\leq R
\leq25\arcmin$. Figure \ref{fmf} shows a break of MFs followed by slope
flattening for masses in the range $m\leq 0.95\,M_{\odot}$, which is more
noticeable in the central region (panel c). MF break has been observed in many
intermediate-age and old open clusters. The presence of the MF break
essentially reflects the effects of the internal dynamics of clusters on the
MFs and/or some fundamental property of the initial mass function (IMF)
associated to different conditions in star formation \citep{bb05}. Solid lines
in Figure \ref{fmf} represent the fitted power-law functions for masses in the
range $0.95\leq m\leq 1.85\,M_{\odot}$. The derived fitting parameters for MFs
of NGC 7789 are listed in Table \ref{tmf}. Figure \ref{fmf} and Table \ref{tmf}
indicate that the MF slopes flatten from the outskirts to the inner regions.
This flat reflects the advanced dynamical state of NGC 7789, particularly the
effects of mass segregation analyzed in the previous section. The overall MF
slope $\alpha=-0.96$ in the mass range $0.95\leq m\leq 1.85\,M_{\odot}$ is much
flatter than the universal IMF slope $\alpha=-2.3$ \citep{kr01}. \citet{ma94}
also find that the faint MS ($V<15.0$) appears rather depopulated of stars and
derive a slope of $-1.3$ for those faint MS stars. The flat overall MF slope
can be accounted for by low-mass stars evaporation resulting from mass
segregation and external effects such as tidal stripping by the Galactic
gravitational field \citep{bb05}. Thus we can't get the IMF of this cluster
from the present observed MF.

To check the effects for our derived MFs of NGC 7789 caused by adopting
different photometric limiting probabilities for choosing candidate member
stars, limiting probability 50\% and 1\% are used to determine candidate member
stars. Table \ref{tmf} lists the fitted parameters for MFs of NGC 7789 with
limiting probabilities 50\% and 1\% respectively. Table \ref{tmf} indicates
that the derived MF slopes in different regions in the cluster are consistent
within uncertainties for different limiting probabilities.
\section{SUMMARY}
In this paper, we present new BATC 13-band photometric results for the
intermediate-age open cluster NGC 7789. Comparing the observed SEDs of cluster
member stars with the theoretical SEDs of Padova 2000 models, we derived a set
of best-fitting fundamental parameters for this cluster: an age of
$\log\,t=9.15\pm0.05$ ($1.4\pm0.1$) Gyr, a distance modulus
$(m-M)_{0}=11.27\pm0.04$, a reddening $E(B-V)=0.28\pm0.02$, and a metallicity
$Z=0.019$. The theoretical SEDs fitted the observed SEDs of RGB stars and MS
stars very well, but can't reproduce the observed SEDs of blue stragglers in
NGC 7789. Our derived fundamental parameters for NGC 7789 are consistent with
recently derived results for this cluster  \citep{gim98b,ba04}.

Surface density profiles for stars within different magnitude (mass) ranges are
fitted by King model \citep{king62} and the core radius $R_{c}$, tidal radius
$R_{t}$ and the concentration parameters $C_{0}=\log\,(R_{t}/R_{c})$ are
derived. For stars within limiting magnitudes of 19.0 in the BATC $e$ band, we
derived $R_{c}=7.52\arcmin$ and $R_{t}=28.84\arcmin$.  The derived
concentration parameters $C_{0}$ are significantly different for stars within
different mass ranges: the $C_{0}$ is the minimum for stars within the
highest-mass range contrast to the value for stars within the lowest-mass
range. This structure parameter variation indicates strong mass segregation in
NGC 7789.

MFs for MS stars with masses from 0.95 to 1.85 $M_{\odot}$ in different spatial
regions of the cluster are fitted by a power-law function $\phi(m)\propto
m^{\alpha}$. We derived $\alpha=-0.96$ for the whole sample of candidate member
stars. The MF slopes significantly change from the central region to the
outskirts: $\alpha=-0.71$ for the central region and $\alpha=-1.20$ for the
outskirts. The significant variation of MF slopes in different spatial regions
also indicates the presence of strong mass segregation and that NGC 7789 has
undertaken strong dynamical evolution.

\acknowledgments This research has made use of the Astrophysical Integrated
Research Environment(AIRE) which is operated by the Center for Astrophysics,
Tsinghua University. We also thank an anonymous referee for a number of
suggestions that improved the rigor and clarity of the paper. This work has
been supported in part by the Chinese National Science Foundation, No.
10473012, 10573020, 10603006, 10673026, 10373020 and 10673012. This research
has made use of the WEBDA database, operated at the Institute for Astronomy of
the University of Vienna.

\clearpage


\begin{deluxetable}{cccccc}
\tabletypesize{\scriptsize} \tablecolumns{6} \tablewidth{0pt}
%\rotate
\tablecaption{Literature Estimates of Fundamental Parameters for NGC
7789\label{told}}
\tablehead{\colhead{$E(B-V)$}&\colhead{$(m-M)_{0}$}&\colhead{[Fe/H]}&\colhead{Age
(Gyr)}&\colhead{Method}&\colhead{References}} \startdata
0.28&$11.36\pm0.2$&\nodata&\nodata&$UBV$ photoelectric and photographic
photometry&1\\
0.23&11.31&\nodata&\nodata&determined from existing data\tablenotemark{a}&2\\
$0.32\pm0.03$&$11.0\pm0.15$&\nodata&\nodata&$uvby\textrm{H}_{\beta}$
photoelectric photometry, 12 blue stragglers&3\\
0.26&$11.5\pm0.2$&$-0.2\pm0.1$&\nodata&$UBViyz$ photoelectric photometry, 19
red
giants&4\\
$0.24\pm0.01$&$11.5\pm0.1$&Solar&\nodata&$UBV$ and DDO photoelectric
photometry,
22 red giants&5\\
$0.22\pm0.02$&\nodata&$-0.35\pm0.07$&\nodata&determined from existing
data\tablenotemark{b}, 12 red giants&6\\
0.27&\nodata&\nodata&\nodata&determined from existing data\tablenotemark{c}, 7
blue stragglers &7\\
$0.31\pm0.03$&$11.34\pm0.3\tablenotemark{*}$&$-0.12\pm0.10\tablenotemark{d}
$&$1.6\pm0.5$&$uvby$ photoelectric photometry, 28 blue stragglers&8\\
\nodata&\nodata&$-0.1\pm0.2$&\nodata&high-resolution CCD spectroscopy, 6
giants&9\\
\nodata&\nodata&-0.05\tablenotemark{e}&\nodata&Washington photoelectric
photometry, 3 red giants&10\\
0.35&11.7&\nodata&1.1&determined from exiting data\tablenotemark{a},
isochrone-fitting&11\\
$0.32\pm0.03$&$11.34\pm0.2\tablenotemark{*}$&Solar&$1.2\pm0.3$&$BV$ CCD
photometry, isochrone-fitting&12\\
\nodata&\nodata&\nodata&$1.34\pm0.18$&$\bigtriangleup V$
method\tablenotemark{a}&13\\
0.29\tablenotemark{f}&11.55\tablenotemark{*}&$-0.08\pm0.02$&\nodata& Determined
from existing data\tablenotemark{b}&14\\
0.28&$\leqslant11.33\tablenotemark{*}$&$-0.2-0.0$&$1.6-1.7$&$VI$ CCD
photometry,
$\sim 18^{'}\times18^{'}$ FOV, isochrone-fitting&15\\
0.30&11.25&$-0.25\pm0.11$&1.4&IR $JK$ CCD photometry, $8^{'}\times8^{'}$ FOV,
isochrone-fitting&16\\
\nodata&\nodata&$-0.16$&\nodata&high-resolution CCD spectroscopy, 8 blue
stragglers&17\\
$0.27\pm0.09$&\nodata&$-0.24\pm0.09$&\nodata&moderate-resolution CCD
spectroscopy, 57 red giants&18\\
$0.25\pm0.02$&$11.32\pm0.03$&$-0.18\pm0.09$&\nodata&Vilnius photoelectric
photometry, 24 red giants&19\\
\nodata&\nodata&$-0.04\pm0.05$&\nodata&high-resolution CCD spectroscopy, 9 red
giants&20\\
\enddata
\tablenotetext{a}{$UBV$ data taken from \citet{bu58}} \tablenotetext{b}{DDO
photoelectric data taken from \citet{ja77}}
\tablenotetext{c}{$uvby\textrm{H}_{\beta}$ data taken from \citet{st70}, $UBV$
data taken from \citet{bu58}} \tablenotetext{d}{Adopting [Fe/H]$=+0.13$ for the
Hyades \citep{bo90}} \tablenotetext{e}{Reported as [A/H] in Washington
photometry system.} \tablenotetext{f}{Averaged result taken from \citet{tw97}.}
\tablenotetext{*}{Inferred from the reported apparent modulus $(m-M)_{V}$,
adopting $A_{V}/E(B-V)=3.1$.} \tablerefs{(1) Burbidge \& Sandage 1958; (2) Arp
1962; (3) Strom \& Strom 1970; (4) Jennens \& Helfer 1975; (5) Janes 1977; (6)
Clari\'{a} 1979; (7) Breger \& Wheeler 1980; (8) Twarog \& Tyson 1985; (9)
Pilachowski 1985; (10) Canterna et al. 1986; (11) Mazzei \& Pigatto 1988; (12)
Martinez Roger et al. 1994; (13) Carraro \& Chiosi 1994; (14) Twarog et al.
1997; (15) Gim et al. 1998b; (16) Vallenari et al. 2000; (17) Sch\"{o}nberner
et al. 2001; (18) Friel et al. 2002; (19) Barta\v{s}i\={u}t\.{e} \&
Tautvai\v{s}ien\.{e} 2004; (20) Tautvai\v{s}ien\.{e} et al. 2005.}
\end{deluxetable}

\begin{deluxetable}{cccccc}
\tablecolumns{6}\tablewidth{0pt} \tablecaption{Parameters of 13 BATC Filters
and Statistics of Our Observations\label{tobs}}
\tablehead{\colhead{No.}&\colhead{Filter
Name}&\colhead{$\lambda_{eff}$}&\colhead{FWHM}
&\colhead{Exposure}&\colhead{Numbers of Images}\\
\colhead{}&\colhead{}&\colhead{(\AA)}&\colhead{(\AA)}&\colhead{(Sec)}
&\colhead{}} \startdata
1&c&4194&309&4200&5\\
2&d&4540&332&16500&16\\
3&e&4925&374&24721&22\\
4&f&5267&344&11400&11\\
5&g&5790&289&7500&7\\
6&h&6074&308&6300&6\\
7&i&6656&491&4080&5\\
8&j&7057&238&6300&6\\
9&k&7546&192&8280&9\\
10&m&8023&255&10200&10\\
11&n&8484&167&4200&5\\
12&o&9182&247&13200&14\\
13&p&9739&275&2400&5\\
\enddata
\end{deluxetable}

\begin{deluxetable}{ccccc}
 \tablecolumns{4}\tablewidth{0pt}
\tablecaption{Completeness Analysis Results for NGC 7789\label{tadds}}
\tablehead{\colhead{$\Delta e$}&\colhead{$0.0<R<25\arcmin$}&\colhead{$0.0<R
<10\arcmin$}&\colhead{$10\arcmin \leq R <25\arcmin$}} \startdata
13 - 14& 0.99&  0.98&  1.00\\
14 - 15& 0.96&  0.96&  0.97\\
15 - 16& 0.94&  0.93&  0.95\\
16 - 17& 0.91&  0.88&  0.93\\
17 - 18& 0.86&  0.81&  0.88\\
18 - 19& 0.74&  0.68&  0.75\\
19 - 20& 0.43&  0.40&  0.44\\
\enddata
\end{deluxetable}

\begin{deluxetable}{cccccccccc}
 \tablecolumns{10}\tablewidth{0pt}
\tablecaption{Fitted Parameters for Surface Density Profiles of NGC
7789\label{tking}} \tablehead{\colhead{Magnitude Range}&\colhead{$R_{c}$
(\arcmin)}&\colhead{$R_{t}$ (\arcmin)}&\colhead{$C_{0}$}&\colhead{$R_{c}$
(\arcmin)}&\colhead{$R_{t}$ (\arcmin)}&\colhead{$C_{0}$}&\colhead{$R_{c}$
(\arcmin)}&\colhead{$R_{t}$ (\arcmin)}&\colhead{$C_{0}$}\\
\colhead{}&\multicolumn{3}{c}{($P_{\textrm{\small{phot}}}>10\%$)}&\multicolumn{3}{c}{($P_{\textrm{\small{phot}}}>50\%$)}&\multicolumn{3}{c}{($P_{\textrm{\small{phot}}}>1\%$)}}
\startdata $e<19.0$&7.52&28.84&0.58&7.39&28.16&0.58&7.69&29.48&0.58\\
$e<15.0$&3.87&40.80&1.02&4.32&27.50&0.80&4.02&33.65&0.92\\
$15.0\leq\,e<17.0$&8.37&26.63&0.50&9.19&22.55&0.39&9.41&23.69&0.40\\
$17.0\leq\,e<19.0$&11.92&27.87&0.37&13.61&26.39&0.29&11.58&31.15&0.43\\
\enddata
\end{deluxetable}

\begin{deluxetable}{cccc}
\tablecolumns{4}\tablewidth{0pt} \tablecaption{Fitted Parameters for Mass
Functions of NGC 7789\label{tmf}}\tabletypesize{\small}
\tablehead{\colhead{Distance $R$ (\arcmin)}&\colhead{$\alpha$
($P_{\textrm{\small{phot}}}>10\%$)}&\colhead{$\alpha$
($P_{\textrm{\small{phot}}}>50\%$)}&\colhead{$\alpha$
($P_{\textrm{\small{phot}}}>1\%$)}}\startdata
$0.0\leq\,R<20.0$&$-0.96$&$-0.90$&$-0.97$\\
$0.0\leq\,R<10.0$&$-0.71$&$-0.65$&$-0.76$\\
$10.0\leq\,R<20.0$&$-1.20$&$-1.21$&$-1.18$\\
\enddata
\end{deluxetable}


\begin{thebibliography}{}
\bibitem[Arp(1962)]{ar62} Arp, H. 1962, \apj, 136, 66
\bibitem[Barta\v{s}i\={u}t\.{e} \& Tautvai\v{s}ien\.{e}(2004)]{ba04}
 Barta\v{s}i\={u}t\.{e}, S., \& Tautvai\v{s}ien\.{e}, G. 2004, \apss, 294, 225
 \bibitem[Bergond et al.(2001)]{be01} Bergond, G., Leon, S., \&
 Guibert, J. 2001, \aap, 377, 462
\bibitem[Boesgaard \& Friel(1990)]{bo90} Boesgaard, A. M., \& Friel, E. D. 1990,
\apj, 351, 467
\bibitem[Bonatto \& Bica(2005)]{bb05} Bonatto, C., \&
Bica, E. 2005, \aap, 437, 483
\bibitem[Bonnell \& Davies(1998)]{bd98} Bonnell, I. A., \& Davies, M.
B. 1998, \mnras, 295, 691
\bibitem[Breger \& Wheeler(1980)]{br80} Breger, M., \& Wheeler, C. 1980, \pasp,
92, 514
\bibitem[Burbidge \& Sandage(1958)]{bu58} Burbidge, E. M., \& Sandage, A. 1956,
\apj, 128, 174
\bibitem[Canterna et al.(1986)]{ca86} Canterna, R., Geisler, D., Harri, H. C.,
Olszewski E., \& Schommer, R. 1986, \aj, 92, 79
\bibitem[Carraro \& Chiosi(1994)]{ca94} Carraro, G., \& Chiosi, C. 1994, \aap,
287, 761
\bibitem[Chen(2000)]{ch00} Chen, A. 2000, Ph.D. thesis, Inst. of Astron., Nat.
Central UNiv., Taiwan
\bibitem[Clari\'{a}(1979)]{cl79} Clari\'{a}, J. J. 1979, \apss, 66, 201
\bibitem[de la Fuente Marcos(1997)]{de97} de la Fuente Marcos, R.
1997, \aap, 322, 764
\bibitem[Fan et al.(1996)]{fan} Fan, X., et al. 1996, \aj, 112, 628
\bibitem[Friel \& Janes(1993)]{fr93} Friel, E. D., \& Janes, K. A. 1993, \aap,
267, 75
\bibitem[Friel et al.(2002)]{fr02} Friel, E. D., Janes, K. A., Tavarez, M., Scot, J., Katsanis, R., Lotz, J., Hong, L., \& Miller, N. 2002, \aj, 124, 2693
\bibitem[Gim et al.(1998a)]{gim98a} Gim, M., Hesser, J. E., McClure, R. D., \&
Stetson, P. B. 1998, \pasp, 110, 1172
\bibitem[Gim et al.(1998b)]{gim98b} Gim, M., VandenBerg, D. A., Stetson, P. B.,
Hesser, J. E., \& Zurek, D. R. 1998, \pasp, 110, 1318
\bibitem[Girardi et al.(2000)]{gir00} Girardi, L., Bressan, A., Bertelli, G., \&
Chiosi, C. 2000, \aaps, 141, 371
\bibitem[Girardi et al.(2000)]{gir00r} Girardi, L., Mermilliod, J. -C., \&
Carraro, G. 2000, \aap, 354, 892
\bibitem[Girardi et al.(2002)]{gir02} Girardi, L., Bertelli, G., Bressan, A., Chiosi, C., Groenewegen, M. A. T., Marigo, P., Salasnich, B., \& Weiss, A. 2002, \aap, 391, 195
\bibitem[Hurley et al.(2005)]{hu05} Hurley, J. R., Pols, O. R.,
Aarseth, S. J., \& Tout, C. A. 2005, \mnras, 363, 293
\bibitem[Janes(1977)]{ja77} Janes, K. A. 1977, \aj, 82, 35
\bibitem[Jennens \& Helfer(1975)]{je75} Jenens, P. A., \& Helfer, H. L. 1975,
\mnras, 172, 681
\bibitem[King(1962)]{king62} King, I. 1962, \aj, 67, 471
\bibitem[Kroupa(2001)]{kr01} Kroupa, P. 2001, \mnras, 322, 231
\bibitem[Martinez Roger et al.(1994)]{ma94} Martinez Roger, M.,  Paez, E.,
Castellani, V., \& Straniero, O. 1994, \aap, 290, 62
\bibitem[Mazzei \& Pigatto(1988)]{ma88} Mazzei, P., \& Pigatto, L. 1988, \aap,
193, 148
\bibitem[McNamara \& Solomon(1981)]{mc81} McNamara, B. J., \& Solomon, S. 1981,
\aaps, 43, 33
\bibitem[Pilachowski(1985)]{pi85} Pilachowski, C. A. 1985, \pasp, 97, 801
\bibitem[Portegies Zwart et al.(2001)]{pz01} Portegies Zwart, S. F.,
Mcmillan, S. L. W., Hut, P., \& Makino, J. 2001, \mnras, 321, 199
\bibitem[Schlegel et al.(1998)]{sch} Schlegel, D. J., Finkbeiner, D. P., \&
Davis, M. 1998, \apj, 500, 525
\bibitem[Sch\"{o}nberner et al.(2001)]{sch01}Sch\"{o}nberner, D., Andrievsky, S.
M., \& Drilling, J. S. 2001, \aap, 366, 490
\bibitem[Stetson(1987)]{stet} Stetson, P. B. 1987, \pasp, 99, 191
\bibitem[Strom \& Strom(1970)]{st70} Strom, K. M., \& Strom, S. E. 1970, \apj,
162, 523
\bibitem[Tautvai\v{s}ien\.{e} et al.(2005)]{ta05} Tautvai\v{s}ien\.{e}, G.,
Edvardsson, B., Puzeras, E., \& Ilyin, I. 2005, \aap, 431, 933
\bibitem[Twarog \& Tyson(1985)]{tw85} Twarog, B. A., \& Tyson, N. 1985, \aj, 90,
124
\bibitem[Twarog et al.(1997)]{tw97} Twarog, B. A., Ashman, K. M., \&
Anthony-Twarog, B. J. 1997, \aj, 114, 2556
\bibitem[Vallenari et al.(2000)]{va00} Vallenari, A., Carraro, G., \& Richichi,
A. 2000, \aap, 353, 147
\bibitem[Wu et al.(2005)]{wu05} Wu, Z. Y., Zhou, X., Ma, J., Jiang, Z. J., \&
Chen, J. S. 2005, \pasp, 117, 32
\bibitem[Wu et al.(2006)]{wu06} Wu, Z. Y., Zhou, X., Ma, J., Jiang, Z. J., \&
Chen, J. S. 2006, \pasp, 118, 1104
\bibitem[Yan et al.(2000)]{yan} Yan, H., et al. 2000, \pasp, 112, 691
\bibitem[Zhou et al.(2001)]{zhou01} Zhou, X., Jiang, Z., Xue, S., Wu, H., Ma, J., \& Chen, J. 2001,
\cjaa, 1, 372
\bibitem[Zhou et al.(2003)]{zhou03} Zhou, X., et al. 2003, \aap, 397, 361
\end{thebibliography}
\end{document}